\documentclass[twocolumn,aps,pra]{revtex4-2}
\usepackage{bm}
\usepackage{amsfonts,amsmath,amssymb,latexsym}
\usepackage{graphicx}
\usepackage{float}
\usepackage{mathtools}
\usepackage{dcolumn}
\usepackage{hyperref}
\begin{document} 
\title{Paraxial Dirac equation}
\author{Tomasz Rado\.zycki}
\email{t.radozycki@uksw.edu.pl}
\affiliation{Faculty of Mathematics and Natural Sciences, College of Sciences, Institute of Physical Sciences, Cardinal Stefan Wyszy\'nski University, W\'oycickiego 1/3, 01-938 Warsaw, Poland} 

\begin{abstract}
In this work, the paraxial approximation of the free Dirac equation is examined. The results are first obtained by constructing superpositions of exact solutions with suitable profiles, which are borrowed from paraxial optics. In this manner, the paraxial Dirac beams are obtained in four cases: as Gaussian, Bessel-Gaussian, modified Bessel-Gaussian and elegant Laguerre-Gaussian beams. In the second part of the work, the paraxial Dirac equation is derived, and then its solutions in the aforementioned cases are directly obtained. All the resulting wave functions conform to those derived formerly by virtue of superpositions, except for terms, that are negligible upon the assumption that the paraxial functions along the propagation axis vary only slightly over a distance equal to the de Broglie wavelength, which is the standard paraxial requirement.
\end{abstract}

\maketitle

\section{Introduction}\label{intr}

The paraxial approximation is the standard practice in laser optics \cite{kl,lax,saleh,sie}, where one deals with beams well collimated along the propagation axis. Within this approximation it is assumed that after a rapidly varying factor of the type of $e^{ipz}$ is extracted, the variation of the wave function over distances of order of one wavelength is minimal. Alternatively one can say that the momenta perpendicular to the main axis are negligible as compared to the longitudinal component. The resulting paraxial equation, which in the scalar form is analogous to (\ref{pareq}), has been widely explored providing rigorous solutions in the form of various beams: Gaussian beams~\cite{kl,saleh,sie,davis,nemo,mw,ibbz,sesh,gustavo,tt,er,selina}, regular Bessel-Gaussian (BG) beams~\cite{saleh,she,gori,april1,mendoza} and modified ones (mBG)~\cite{bagini} as well as Laguerre-Gaussian (LG) beams~\cite{sie,saleh,mendoza,lg,lg2,april2,april3,nas} or Kummer-Gaussian (KG) (i.e.,  Hypergeometric-Gaussian) beams~\cite{kot,karimi}. All these cylindrical beams have vortex-like character and are endowed with the orbital angular momentum (OAM). Among non-cylindrical beams one can mention Hermite-Gaussian ones~\cite{kl,sie} or certain more general paraxial beams~\cite{trgen}. These radiation modes have found numerous applications in physics, optical technologies, and even biology and medicine~\cite{ste,fazal,pad,woe,bowpa,grier1,kol,alt,nis,cc}.

Compared to this, substantially less attention (which does not mean little) has been paid to paraxial electron beams. This is obviously due to the fact that the prediction of vortex solutions in the propagating electron wave packets was first done -- for the Schr\"odinger equation -- only half a century after optical discoveries \cite{bbsn}. At least from the theoretical point of view, owing to the similarity between the Helmholtz equation and the stationary Schr\"odinger equation, the optical beams mentioned above should find their counterparts in the propagation of electrons. When it comes to the experimental generation of the latter, it involves more challenges because of the remarkably small, as compared to that in optics, de Broglie wavelength. Nevertheless, in a short time, such vortex packets were experimentally generated \cite{uchida,ver,mm,sht,ssv}, recently reaching the extremely high values of OAM \cite{tav}. 

The natural framework to study the intrinsic OAM of the electron beam is, however, not the Schr\"odinger but the Dirac equation. The first vortex-like solution of this equation for freely moving electrons in the form of the nondifractive Bessel beam was obtained in \cite{bdn}, followed by those in various external fields \cite{karl,hay,raj,zou,vE,camp}.

The structured electron beams, especially those endowed with OAM, can find numerous applications in electron microscopy \cite{too} and in scattering experiments on various atomic and subatomic systems \cite{yu,ser,kks,ed}. The properties of a twisted beam are significantly different from those attributable to a free plane-wave electron: in particular, its magnetic moment is associated not only with spin but also with OAM, so it can acquire much larger values, and constitutes a stronger source of the radiative emission \cite{add,lar,iva}. Thanks to this property it can serve as a useful tool to probe the magnetic media \cite{ru,yu}.

Up to our knowledge significantly less focus has been directed to the paraxial approximation of the Dirac equation. The main tool seems here to be so far the Foldy-Wouthuysen (FW) transformation \cite{folw}, which decouples the large and small components of the bispinor from each other and then boils down the problem of deriving the paraxial solution to that known from the scalar optics\cite{bdn,barn17}. 

In the present work our intention is to propose the explicit form of the paraxial Dirac equation which allows for direct calculations. Such an equation is suggested in Section \ref{peq} and then a couple of solutions to this equation are obtained: these are Gaussian, Bessel-Gaussian, modified Bessel-Gaussian and elegant Laguerre-Gaussian beams. Earlier, is Section \ref{potb} the various approximations of the exact solution of the standard Dirac equation in cylindrical coordinates leading to the same four paraxial beams are found. They are obtained by superposing exact solutions with some properly selected amplitudes. The results of both sections turn out to be mutually consistent within the applied paraxial approximation. 

Throughout this work the system of units for which $c=\hbar=1$ is used, and the signature of the Minkowski metric tensor $\eta$ is chosen as $(+---)$. If so, the co- and contravariant four-vectors satisfy  
\begin{equation}\label{degd}
V^\mu=\eta^{\mu\nu}V_\nu, 
\end{equation}
and hence
\begin{equation}\label{degda}
 V^0=V_0,\; V^k=-V_k \;\;(k=1,2,3).
\end{equation}
The free Dirac equation has the form
\begin{equation}\label{diraceq}
\left(i\gamma^\mu\partial_\mu -m\right)\Psi(t,\mathbf{r})=0,
\end{equation}
where for $\gamma$ matrices the Dirac representation is chosen. 
Substituting
\begin{equation}\label{elfi}
\Psi(t,\mathbf{r})=e^{-iEt}\psi(\mathbf{r}).
\end{equation}
we get the stationary equation
\begin{equation}\label{diraceq2}
\left(\gamma^0E+i\bm{\gamma}\bm{\partial} -m\right)\psi(\mathbf{r})=0,
\end{equation}
which will be the concern of this paper. As mentioned above, we are mainly interested in cylindrical beams, and therefore, henceforth the cylindrical coordinates are used, in which Eq. (\ref{diraceq2}) reads
\begin{equation}\label{diraccy}
\left(\gamma^0E+i\gamma^\rho\partial_\rho+\frac{i}{\rho}\,\gamma^\varphi\partial_\varphi+i\gamma^3\partial_z
-m\right)\psi(\rho,\varphi,z)=0
\end{equation}
where $\partial_\rho=\partial/\partial_\rho$ etc., and
\begin{equation}\label{gamr}
\gamma^\rho=\left(\begin{array}{cc} 0 &\sigma_\rho\\ -\sigma_\rho&0\end{array}\right), \quad 
\gamma^\varphi=\left(\begin{array}{cc} 0 &\sigma_\varphi\\ -\sigma_\varphi&0\end{array}\right)
\end{equation}
with
\begin{equation}\label{sigr}
\sigma_\rho=\left(\begin{array}{cc} 0 &e^{-i\varphi}\\ e^{i\varphi}&0\end{array}\right), \quad 
\sigma_\varphi=\left(\begin{array}{cc} 0 &-ie^{-i\varphi}\\ ie^{i\varphi}&0\end{array}\right)
\end{equation}

\section{Paraxial approximations of the exact beam}\label{potb}

\subsection{The exact cylindrical solution }\label{cyso}

In order to find the monoenergetic solutions of the Dirac equation (\ref{diraccy}) the conventional practice is the use of the substitution
\begin{equation}\label{bie}
\psi(\rho,\varphi,z)=\left[\begin{array}{c}\phi(\rho,\varphi,z)\\ \chi(\rho,\varphi,z)\end{array}\right],
\end{equation}
where $\phi$ and $\chi$ are two-component spinors. Equation (\ref{diraccy}) can then be rewritten in the standard way in the form of two coupled equations
\begin{subequations}\label{gd}
\begin{align}
&(E-m)\phi=-i\left(\sigma_\rho\partial_\rho+\frac{1}{\rho}\, \sigma_\phi\partial_\phi+\sigma_z\partial_z\right)\chi,\label{gdu}\\
&(E+m)\chi=-i\left(\sigma_\rho\partial_\rho+\frac{1}{\rho}\, \sigma_\phi\partial_\phi+\sigma_z\partial_z\right)\phi,\label{gdd}
\end{align}
\end{subequations}
which can then be decoupled at the price of producing the equation of the second degree: 
\begin{equation}\label{phis}
(E^2-m^2)\phi=-\left(\partial_\rho^2+\frac{1}{\rho}\,\partial_\rho+\frac{1}{\rho^2}\,\partial_\varphi^2+\partial_z^2\right)\phi,
\end{equation}
and identically for lower bispinor components.

For cylindrical solutions, we are interested in, the spinor $\phi$ exhibits the $\varphi$-dependence in the form of: 
\begin{equation}\label{fgbie}
\phi_n(\rho,\varphi,z)=\left[\begin{array}{c}f_n(\rho,\varphi,z)\\ g_n(\rho,\varphi,z)\end{array}\right]=e^{in\varphi}\left[\begin{array}{c}\mathrm{f}_n(\rho,z)\\ \mathrm{g}_n(\rho,z)\end{array}\right]
\end{equation}
The parameter $n$ assumes the integral values and accounts for the orbital angular momentum of the beam~\cite{blirep}. 
The equations satisfied by the upper and lower components of $\phi_n$ are identical:
\begin{equation}\label{fgeq}
\left(\partial_\rho^2+\frac{1}{\rho}\,\partial_\rho-\frac{n^2}{\rho^2}+\partial_z^2+p^2\right)\left\{\begin{array}{c}\mathrm{f}_n\\ \mathrm{g}_n\end{array}\right\}=0,
\end{equation}
where $p^2=E^2-m^2$. The elimination of the $z$ dependence by extracting the factor $e^{ip_z z}$, visibly leads to the Bessel equation~\cite{gr} in variable $\rho$. Consequently, the cylindrical solution of (\ref{phis}) is expressed in terms of the function
\begin{equation}\label{fj}
f_n(\rho,\varphi,z)=c_ne^{ip_z z}e^{in\varphi}J_n(p_\rho \rho),
\end{equation}
up to a normalisation constant $c_n$, where the radial momentum component is defined by $p_\rho^2=p^2-p_z^2$. Obviously the identical formula applies to the function $g_n(\rho,\varphi,z)$. 

Now, from Eq. (\ref{gdd}) it follows that
\begin{equation}\label{fgd}
\chi_n(\rho,\varphi,z)=\frac{-i}{E+m}\left[\begin{array}{c}e^{-i\varphi}(\partial_\rho g_n+\frac{n}{\rho}\,g_n)+\partial_z f_n\\ e^{i\varphi}(\partial_\rho f_n-\frac{n}{\rho}\,f_n)-\partial_z g_n\end{array}\right].
\end{equation}
However, we will not deal with solutions that are superpositions of different spin states, and thus the use of two different functions $f_n$ and $g_n$ satisfying identical differential equations is unnecessary. For the description of definite spin states, only one of them (denoted henceforth as $f_n$) -- once acting as a function $f_n$ and once as $g_n$ -- is adequate.

Consequently the followig formulas for``spin-up'' 
\begin{equation}\label{fubie}
\psi_{n\uparrow}(\rho,\varphi,z)=e^{in\varphi}\left[\begin{array}{c}f_n(\rho,\varphi,z)\\ 0\\ \frac{1}{E+m}\left(\begin{array}{c}-i\partial_z f_n(\rho,\varphi,z)\\ ip_\rho f_{n+1}(\rho,\varphi,z)\end{array}\right)\end{array}\right],
\end{equation}
and ``spin-down'' solutions
\begin{equation}\label{fubid}
\psi_{n\downarrow}(\rho,\varphi,z)=e^{in\varphi}\left[\begin{array}{c}0\\ f_n(\rho,\varphi,z)\\ \frac{1}{E+m}\left(\begin{array}{c}-ip_\rho f_{n-1}(\rho,\varphi,z)\\ i\partial_z f_n(\rho,\varphi,z)\end{array}\right)\end{array}\right].
\end{equation}
are obtained (identical to those of \cite{bdn,blirep}, up to the normalization constants), where the following identities for the Bessel functions of the first kind~\cite{gr} have been made use of
\begin{subequations}\label{ide}
\begin{align}
&x\,\frac{d}{dx}J_n(x)-nJ_n(x)=-xJ_{n+1}(x),\label{ide1}\\
&x\,\frac{d}{dx}J_n(x)+nJ_n(x)=xJ_{n-1}(x),\label{ide2}
\end{align}
\end{subequations}
in order to simplify the expression (\ref{fgd}). These are the simplest solutions possessing orbital angular momentum. Note, however, that the value of OAM for the cylindrical solutions of the Dirac equation even characterized by the concrete value of $n$ is not well defined, since these expressions contain admixtures of angular momentum equal to $(n\pm 1)\hbar$ in the lower components. The occurrence of these terms is an implication of the spin-orbit interaction and means that the bispinors do not represent the eigenstates individually of OAM and spin.

These exact modes (\ref{fubie}) and (\ref{fubid}) are non-normalizable due to their infinite spatial extent in the variable $\rho$. Real waves are obviously spatially limited. In the following sections these expressions will, however, provide the starting point for performing the paraxial approximations leading to various beams with Gaussian profile in the perpendicular plane (and thereby allowing them to be normalized on this plane) and delocalized in longitudinal direction. These paraxial solutions are obtained from the above exact ones through the following four steps:
\begin{enumerate}
\item First it is assumed that the beam is well collimated along the $z$-axis, and hence the radial component of the momentum is small, i.e., $p_\rho\ll p$. This allows to make use of the approximation common in optics~\cite{saleh}:
\begin{equation}\label{pza}
e^{ip_zz}=e^{i\sqrt{p^2-p_\rho^2}}\approx e^{ipz}e^{-i\frac{p_\rho^2}{2p}\,z}.
\end{equation}
\item Second, the rapidly oscillating factor $e^{ipz}$ can be extracted leaving the slowly varying functions labeled with a ``hat'':
\begin{equation}\label{deha}
F=e^{ipz}\widehat{F},\;\;\mathrm{and}\;\; i\partial_z F=e^{ipz}(-p\widehat{F}+i\partial_z \widehat{F}),
\end{equation}
with $F$ standing for $\psi_n$, $\phi_n$, $\chi_n$, $f_n$, $g_n$ and so on. The second exponential factor in the approximation (\ref{pza}) is retained in the definition of $\widehat{f}_n$, i.e.,
\begin{equation}\label{fjh}
\widehat{f}_n(\rho,\varphi,z)=c_ne^{in\varphi}e^{-i\frac{p_\rho^2}{2p}\,z}J_n(p_\rho \rho).
\end{equation}
\item Next, the ``hatted'' wavefunction $\widehat{\psi}_n$ is subject to some integral transform with a universial Gaussian factor
\begin{equation}\label{intt}
\widetilde{\psi}_n(\rho,\varphi,z)=\int\limits_0^\infty d p_\rho p_\rho A(p_\rho)e^{-\frac{w_0^2p_\rho^2}{4}}\widehat{\psi}_n(\rho,\varphi,z),
\end{equation}
and various choices of the prefactor $A(p_\rho)$, which lead to different classes of (Gaussian) paraxial beams.
\item Finally, the $z$ derivative of the formfactor $\widetilde{f}_n(\rho,\varphi,z)$ which is introduced below, can be neglected as compared to the value of  momentum, or more precisely:
\begin{equation}\label{apl}
|\partial_z \widetilde{f}_n(\rho,\varphi,z)|\ll |p \widetilde{f}_n(\rho,\varphi,z)|.
\end{equation}
This is the obvious implication of the assumption that the value of the function $\widetilde{f}_n$ vary only slightly along the propagation axis over a distance of the order of the electron's de Broglie wavelength $\lambda_{\mathrm{dB}}$. 
\end{enumerate}
The aforementioned procedure appears to be preferable to the oversimplified one suggested in \cite{bdn,lloyd,blirep}, which would, among others, lose the Gaussian factors. The nontrivial paraxial Dirac beams are obtained not only by approximating a pure Bessel beam, i.e. ignoring the transverse components of momentum, but also through appropriate superposition of waves. In the following, four approximations outlined above leading to four different paraxial beams will be examined in turn.

\subsection{Gaussian beam}\label{gaud}

The Gaussian beam is obtained by inserting into(\ref{intt}) the prefactor $A(p_\rho)$ in the form
\begin{equation}\label{aga}
A(p_\rho)=p_\rho^n,
\end{equation}
and performing the following integral~\cite{trhan}:
\begin{equation}\label{prp}
\widetilde{\psi}_n(\rho,\varphi,z)=\int\limits_0^\infty d p_\rho p_\rho^{n+1}e^{-\frac{w_0^2p_\rho^2}{4}}\,\widehat{\psi}_n(\rho,\varphi,z),
\end{equation}
which will be henceforth called the ``Gaussian Paraxial Transform'' (GPT) and denoted with $G_n[\widehat{\psi}]$. It represents some specific superposition of exact modes if the value of $w_0$ is large enough to legitimize the use of the approximation (\ref{pza}), which in practice indicates $w_0\gg \lambda_{\mathrm{dB}}$.

In order to derive the paraxial approximation of (\ref{fubie}) and (\ref{fubid}) the following GPTs of the ``hatted'' quantities are needed. First
\begin{eqnarray}
G_n[\widehat{f}_n(\rho,\varphi,z)]&=&e^{in\varphi}\int\limits_0^\infty dp_\rho\, p_\rho^{n+1}e^{-\frac{\alpha(z)p_\rho^2}{4}}J_n(p_\rho\rho)\label{inta}\\
&=&e^{in\varphi}\left(\frac{2}{\alpha(z)}\right)^{n+1}\!\!\!\!\rho^ne^{-\frac{\rho^2}{\alpha(z)}}=:\widetilde{f}_n(\rho,\varphi,z)
\nonumber
\end{eqnarray}
where $\alpha(z)=w_0^2+\frac{2iz}{p}$ is the complex beam parameter known from optics, and then 
\begin{subequations}\label{tra}
\begin{align}
&G_n[-i\partial_z\widehat{f}_n(\rho,\varphi,z)]=-i\partial_z\widetilde{f}_n(\rho,\varphi,z),\label{tra1}\\
&G_n[ip_\rho\widehat{f}_{n+1}(\rho,\varphi,z)]=\widetilde{f}_{n+1}(\rho,\varphi,z),\label{tra2}\\
&G_n[-ip_\rho\widehat{f}_{n-1}(\rho,\varphi,z)]=2p\partial_z\widetilde{f}_{n-1}(\rho,\varphi,z).\label{tra4}
\end{align}
\end{subequations}

These results can be collected to yield the paraxial beams describing the cylindrical ``spin-up'' and ``spin-down'' solutions:
\begin{equation}\label{gapu}
\widetilde{\psi}_{n\uparrow}(\rho,\varphi,z)= \left[\begin{array}{c}\widetilde{f}_n\\ 0\\ \frac{1}{E+m}\left(\begin{array}{c}p\widetilde{f}_n-i\partial_z\widetilde{f}_n\\ i\widetilde{f}_{n+1}\end{array}\right)\end{array}\right],
\end{equation}
\begin{equation}\label{gapd}
\widetilde{\psi}_{n\downarrow}(\rho,\varphi,z)= \left[\begin{array}{c}0\\ \widetilde{f}_n\\ \frac{1}{E+m}\left(\begin{array}{c}2p\partial_z\widetilde{f}_{n-1}\\ -p\widetilde{f}_n+i\partial_z\widetilde{f}_n \end{array}\right)\end{array}\right].
\end{equation}
As already mentioned the partial derivative $\partial_z\widetilde{f}_n$ may be omitted relative to $p\widetilde{f}_n$, since by assumption $\widetilde{f}_n$ is a slowly varying function of $z$. As can be seen, even for a paraxial beam with a definite index n, the OAM is not well-defined due to the lower components. The same refers to the following beams.

\subsection{Bessel-Gaussian beam}\label{begd}

Other paraxial solutions can be obtained in an analogous way but with the modification involving the prefactor $A(p_\rho)$ of the paraxial transform (\ref{intt}). In the case of the Bessel-Gaussian beam it takes the form~\cite{trhan}
\begin{equation}\label{agb}
A(p_\rho)=I_n(\chi p_\rho),
\end{equation}
where $\chi$ is a certain parameter related to the aperture angle of the beam~\cite{bor,mad} and $I_n$ stands for the hyperbolic Bessel fuction. Accordingly 
\begin{equation}\label{prpb}
\widetilde{\psi}_n(\rho,\varphi,z)=\int\limits_0^\infty d p_\rho p_\rho I_n(\chi p_\rho)e^{-\frac{w_0^2p_\rho^2}{4}}\,\widehat{\psi}_n(\rho,\varphi,z),
\end{equation}
which might be called the ``Bessel-Gaussian Paraxial Transform'' (BGPT) and is denoted below with $BG_n[\widehat{\psi}]$. 

Without going into calculational details, let us summarize the results of the transformations of all the components necessary to compose the paraxial solutions that exhibit the character of a Bessel-Gauss beam, viz
\begin{subequations}\label{tro}
\begin{align}
&BG_n[\widehat{f}_n(\rho,\varphi,z)]\label{troa1}\\
&\hspace{3ex}=e^{in\varphi}\int\limits_0^\infty dp_\rho\, p_\rho I_n(\chi p_\rho)e^{-\frac{\alpha(z)p_\rho^2}{4}}\,J_n(p_\rho\rho)\nonumber\\
&\hspace{3ex}=e^{in\varphi}\frac{2}{\alpha(z)}e^{\frac{\chi^2-\rho^2}{\alpha(z)}}J_n\left(\frac{2\chi\rho}{\alpha(z)}\right)=:\widetilde{f}_n(\rho,\varphi,z),\label{troa}\\
&BG_n[-i\partial_z\widehat{f}_n(\rho,\varphi,z)]=-i\partial_z\widetilde{f}_n(\rho,\varphi,z),\nonumber\\
&BG_n[ip_\rho\widehat{f}_{n+1}(\rho,\varphi,z)]=\frac{2i\chi}{\alpha(z)}\,\widetilde{f}_{n+1}(\rho,\varphi,z)\label{troa2}\\
&\hspace{23ex}+e^{i\varphi}\frac{2i\rho}{\alpha(z)}\,\widetilde{f}_n(\rho,\varphi,z),\nonumber\\
&BG_n[-ip_\rho\widehat{f}_{n-1}(\rho,\varphi,z)]=-\frac{2i\chi}{\alpha(z)}\,\widetilde{f}_{n-1}(\rho,\varphi,z)\label{troa4}\\
&\hspace{25ex}+e^{-i\varphi}\frac{2i\rho}{\alpha(z)}\,\widetilde{f}_n(\rho,\varphi,z).\nonumber
\end{align}
\end{subequations}
Consequently one obtains
\begin{equation}\label{bgapu}
\widetilde{\psi}_{n\uparrow}(\rho,\varphi,z)= \left[\begin{array}{c}\widetilde{f}_n\\ 0\\ \frac{1}{E+m}\left(\begin{array}{c}p\widetilde{f}_n-i\partial_z\widetilde{f}_n\\ \frac{2i\chi}{\alpha(z)}\,\widetilde{f}_{n+1}+e^{i\varphi}\frac{2i\rho}{\alpha(z)}\,\widetilde{f}_n\end{array}\right)\end{array}\right]
\end{equation}
and 
\begin{equation}\label{bgapd}
\widetilde{\psi}_{n\downarrow}(\rho,\varphi,z)= \left[\begin{array}{c}0\\ \widetilde{f}_n\\ \frac{1}{E+m}\left(\begin{array}{c} -\frac{2i\chi}{\alpha(z)}\,\widetilde{f}_{n-1}+e^{-i\varphi}\frac{2i\rho}{\alpha(z)}\,\widetilde{f}_n\\ -p\widetilde{f}_n+i\partial_z\widetilde{f}_n\end{array}\right)\end{array}\right].
\end{equation}

Again, within the framework of the approximation used, the derivative with respect to $z$ in the lower components can be disregarded in comparison with $p\widetilde{f}_n$.

\subsection{Modified Bessel-Gaussian beam}\label{mbegd}

In order to get the paraxial solution that would correspond to the modified Bessel-Gaussian optical beam, the prefactor in the form of the Bessel function of the first kind has to be used \cite{trhan}
\begin{equation}\label{agba}
A(p_\rho)=J_n(\chi p_\rho).
\end{equation}
The paraxial wavefunction is then obtained by the following integral transform
\begin{equation}\label{prpbm}
\widetilde{\psi}_n(\rho,\varphi,z)=\int\limits_0^\infty d p_\rho p_\rho J_n(\chi p_\rho)e^{-\frac{w_0^2p_\rho^2}{4}}\,\widehat{\psi}_n(\rho,\varphi,z),
\end{equation}
where $\chi$ is again a parameter. This integral can be termed as ``modified Bessel-Gaussian Paraxial Transform'' (mBGPT) and denoted with $mBG_n[\widehat{\psi}]$. The construction of the beam components proceeds similarly to that of the previous section:
\begin{subequations}\label{trom}
\begin{align}
&mBG_n[\widehat{f}_n(\rho,\varphi,z)]\label{troma}\\
&\hspace{3ex}=e^{in\varphi}\int\limits_0^\infty dp_\rho\, p_\rho J_n(\chi p_\rho)e^{-\frac{\alpha(z)p_\rho^2}{4}}\,J_n(p_\rho\rho)\nonumber\\
&\hspace{3ex}=e^{in\varphi}\frac{2}{\alpha(z)}e^{-\frac{\chi^2+\rho^2}{\alpha(z)}}I_n\left(\frac{2\chi\rho}{\alpha(z)}\right)=:\widetilde{f}_n(\rho,\varphi,z),\nonumber\\
&mBG_n[-i\partial_z\widehat{f}_n(\rho,\varphi,z)]=-i\partial_z\widetilde{f}_n(\rho,\varphi,z),\label{troma1}\\
&mBG_n[ip_\rho\widehat{f}_{n+1}(\rho,\varphi,z)]=-\frac{2i\chi}{\alpha(z)}\,\widetilde{f}_{n+1}(\rho,\varphi,z)\label{troma2}\\
&\hspace{23ex}+e^{i\varphi}\frac{2i\rho}{\alpha(z)}\,\widetilde{f}_n(\rho,\varphi,z),\nonumber\\
&mBG_n[-ip_\rho\widehat{f}_{n-1}(\rho,\varphi,z)]=-\frac{2i\chi}{\alpha(z)}\,\widetilde{f}_{n-1}(\rho,\varphi,z)\label{troma4}\\
&\hspace{25ex}+e^{-i\varphi}\frac{2i\rho}{\alpha(z)}\,\widetilde{f}_n(\rho,\varphi,z).\nonumber
\end{align}
\end{subequations}
The obtained form of the bispinors is formally very similar to (\ref{bgapu}) and (\ref{bgapd}), i.e.,
\begin{equation}\label{mbgapu}
\widetilde{\psi}_{n\uparrow}(\rho,\varphi,z)= \left[\begin{array}{c}\widetilde{f}_n\\ 0\\ \frac{1}{E+m}\left(\begin{array}{c}p\widetilde{f}_n-i\partial_z\widetilde{f}_n\\ -\frac{2i\chi}{\alpha(z)}\,\widetilde{f}_{n+1}+e^{i\varphi}\frac{2i\rho}{\alpha(z)}\,\widetilde{f}_n\end{array}\right)\end{array}\right]
\end{equation}
and 
\begin{equation}\label{mbgapd}
\widetilde{\psi}_{n\downarrow}(\rho,\varphi,z)= \left[\begin{array}{c}0\\ \widetilde{f}_n\\ \frac{1}{E+m}\left(\begin{array}{c} -\frac{2i\chi}{\alpha(z)}\,\widetilde{f}_{n-1}+e^{-i\varphi}\frac{2i\rho}{\alpha(z)}\,\widetilde{f}_n\\ -p\widetilde{f}_n+i\partial_z\widetilde{f}_n\end{array}\right)\end{array}\right],
\end{equation}
but it should be remembered that the formfactor $\widetilde{f}_n(\rho,\varphi,z)$ is now represented by means of the formula (\ref{troma}) and the role of the parameter $\chi$ is different \cite{trsup}.

\subsection{Elegant Laguerre-Gaussian beam}\label{elegd}

The last beam to be addressed in this work is that known in optics as (elegant) Laguerre-Gaussian beam. In this case one has to choose the prefactor in the form \cite{trhan}
\begin{equation}\label{prelg}
A(p_\rho)=p_\rho^{n+2q}, 
\end{equation}
where $q$ is a natural number. This choice  leads to the following transform of the wavefunction:
\begin{equation}\label{prpelm}
\widetilde{\psi}_{n,q}(\rho,\varphi,z)=\int\limits_0^\infty d p_\rho p_\rho^{n+2q+1} e^{-\frac{w_0^2p_\rho^2}{4}}\,\widehat{\psi}_n(\rho,\varphi,z),
\end{equation}
 The integral~(\ref{prpelm}) will be referred to below as ``elegant Laguerre-Gaussian Paraxial Transform'' (eLGPT) and denoted in the subsequent formulas with $eLG_{n,q}[\widehat{\psi}]$. Like with the preceding beams, the following building blocks need to be found:
 \begin{subequations}
\begin{align}
&eLG_{n,q}[\widehat{f}_n(\rho,\varphi,z)]\label{trolga}\\
&\hspace{3ex}=e^{in\varphi}\int\limits_0^\infty dp_\rho\, p_\rho^{n+2q+1} e^{-\frac{\alpha(z)p_\rho^2}{4}}\,J_n(p_\rho\rho)\nonumber\\
&\hspace{3ex}=2^qq!\left(\frac{2}{\alpha(z)}\right)^{n+q+1}\!\!\!\!\!\!e^{in\varphi}\rho^ne^{-\frac{\rho^2}{\alpha(z)}}L_q^{(n)}\left(\frac{\rho^2}{\alpha(z)}\right)\\
&\hspace{3ex}=\widetilde{f}_{n,q}(\rho,\varphi,z),\nonumber\\
&eLG_{n,q}[-i\partial_z\widehat{f}_n(\rho,\varphi,z)]=-i\partial_z\widetilde{f}_{n,q}(\rho,\varphi,z),\label{trolga1}\\
&eLG_{n,q}[ip_\rho\widehat{f}_{n+1}(\rho,\varphi,z)]=i\widetilde{f}_{n+1,q}(\rho,\varphi,z)\label{trolga4}\\
&eLG_{n,q}[-ip_\rho\widehat{f}_{n-1}(\rho,\varphi,z)]=-i\widetilde{f}_{n-1,q+1}(\rho,\varphi,z),\nonumber
\end{align}
\end{subequations}
where $L_q^{(n)}(x)$ stand for associated Laguerre polynomials \cite{gr}.
These results allow to write the explicit form of the bispinors corresponding to both spin states as
\begin{equation}\label{elgapu}
\widetilde{\psi}_{n,q\uparrow}(\rho,\varphi,z)= \left[\begin{array}{c}\widetilde{f}_{n,q}\\ 0\\ \frac{1}{E+m}\left(\begin{array}{c}p\widetilde{f}_{n,q}-i\partial_z\widetilde{f}_{n,q}\\ i\widetilde{f}_{n+1,q}\end{array}\right)\end{array}\right]
\end{equation}
and 
\begin{equation}\label{elgapd}
\widetilde{\psi}_{n,q\downarrow}(\rho,\varphi,z)= \left[\begin{array}{c}0 \\ \widetilde{f}_{n,q}\\ \frac{1}{E+m}\left(\begin{array}{c}-i\widetilde{f}_{n-1,q+1}\\ -p\widetilde{f}_{n,q}+i\partial_z\widetilde{f}_{n,q}\end{array}\right)\end{array}\right].
\end{equation}

In the following section we are going to formulate the paraxial Dirac equation, and then to determine its solutions that describe the four beams discussed above. Then they will be compared to those obtained from the superpositions.

\section{Paraxial equation}\label{peq}

\subsection{Derivation of the equation}\label{der}

In order to derive the paraxial Dirac equation the rapidly oscillating $z$-dependence has to be detached from the slow one, as it is usually done in paraxial optics. To this goal, one first factors out the exponential $e^{ikz}$, and then neglects the remaining second derivative $\partial_z^2$ operating on a slowly-varying ``envelope'', which is small compared to $k\partial _z$. In a systematic way this procedure was studied in~\cite{lax75}. 

In the case of the Dirac equation the same result can be achieved by isolating the factor $e^{ipz}$, i.e. by writing
\begin{equation}\label{tpsi}
\psi(\mathbf{r})=e^{ipz}\widehat{\psi}(\mathbf{r}),
\end{equation}
in which case 
\begin{equation}\label{diraceq2a}
\left(\gamma^0E+i\gamma^1\partial_x+i\gamma^2\partial_y+i\gamma^3\partial_z-\gamma^3p -m\right)\widehat{\psi}(\mathbf{r})=0.
\end{equation}
and then introducing a certain matrix $\Gamma$, accompanying the remaining $z$-derivative, in place of $\gamma^3$. The form of this matrix is to be established. One then obtains the paraxial equation in the form of
\begin{equation}\label{diraceq2p}
\left(\gamma^0E+i\gamma^1\partial_x+i\gamma^2\partial_y+i\Gamma\partial_z-\gamma^3p -m\right)\widetilde{\psi}(\mathbf{r})=0,
\end{equation}
and its solutions -- i.e. the paraxial beams -- are marked with a tilda in correspondence to the previous section.

The matrix $\Gamma$ can be expanded in terms of $16$ linearly independent Dirac matrices~\cite{iz} as follows
\begin{equation}\label{Gamma}
\Gamma=S\openone+V_\mu\gamma^\mu+T_{\mu\nu}\sigma^{\mu\nu}+A_\mu\gamma^5\gamma^\mu+B\gamma^5,
\end{equation}
where $\openone$ denotes the unit $4\times 4$ matrix, and the coefficient functions $S,V_\mu,T_{\mu\nu},A_\mu,B$ are to be found. The tensor $T_{\mu\nu}$ is obviously antisymmetric.

We expect each component $\beta=1,\ldots,4$ of the bispinor $\widetilde{\psi}$ to satisfy the standard scalar paraxial equation
\begin{equation}\label{pareq}
\left(\Delta_\perp+2ip\partial_z\right)\widetilde{\psi}_\beta=0.
\end{equation}
This requirement entails the following conditions to be satisfied, which arise upon squaring the operator in (\ref{diraceq2p}):
\begin{align}\label{conds}
\begin{split}
&\Gamma^2=0,\quad \{\Gamma, \gamma^0\}=0,\quad \{\Gamma, \gamma^1\}=0,\\
&\{\Gamma, \gamma^2\}=0,\quad \{\Gamma, \gamma^3\}=-2.
\end{split}
\end{align}

First demanding $\{\Gamma, \gamma^0\}=0$, and exploiting the standard (anti) commutation relations between Dirac matrices, one finds
\begin{equation}\label{co1}
2S\gamma^0+2V^0\openone+2i\left(T^0_{\;\mu}\gamma^\mu-T_{\mu}^{\;0}\gamma^\mu\right)-2A^0\gamma^5=0
\end{equation}
Since matrices $\gamma^0$, $\openone$, $\gamma^\mu$ and $\gamma^5$ are independent, Eq.~(\ref{co1}) implies that
\begin{equation}\label{fico}
 S=0,\qquad V^0=0,\qquad A^0=0,\qquad T^{0k}=-T^{k0}=0,
\end{equation}
and the matrix $\Gamma$ gets reduced to the form
\begin{equation}\label{Gamma1}
\Gamma=-V^k\gamma^k+T^{kj}\sigma^{kj}-A^k\gamma^5\gamma^k+B\gamma^5.
\end{equation}

If one now makes use of the condition $\{\Gamma, \gamma^k\}=0$ for  $k=1,2$, the following equation is obtained (here $j=1,2$ as well):
\begin{equation}\label{co2}
V^k\openone+2iT^{kj}\gamma^j+A^k\gamma^5=0,
\end{equation}
which in turn entails
\begin{equation}\label{fica}
V^k=0,\qquad T^{kj}=-T^{jk}=0,\qquad A^k=0.
\end{equation}
Hence, expression (\ref{Gamma1}) is further simplified to
\begin{equation}\label{Gamma2}
\Gamma=-V^3\gamma^3+T^{3j}\sigma^{3j}+T^{j3}\sigma^{j3}-A^3\gamma^5\gamma^3+B\gamma^5.
\end{equation}
From the last condition of~(\ref{conds}) it stems that
\begin{equation}\label{co3}
V^3\openone-2iT^{3j}\gamma^j-A^3\gamma^5=-\openone,
\end{equation}
and consequently we find
\begin{equation}\label{ficb}
V^3=-1,\qquad T^{3j}=-T^{j3}=0,\qquad A^3=0.
\end{equation}
Finally one gets the concluding form of the matrix $\Gamma$:
\begin{equation}\label{Gamma3}
\Gamma=\gamma^3+B\gamma^5.
\end{equation}
The only unspecified coefficient B can be fixed owing to the nilpotency property (of order $2$) of the matrix $\Gamma$. It is then elementary to infer that the constant $B$, henceforth denoted with $\varepsilon$, equals $\pm 1$, i.e., 
\begin{equation}\label{Gamf}
\Gamma_\varepsilon=\gamma^3+\varepsilon \gamma^5,\quad \varepsilon=\pm 1.
\end{equation}
In particular, in the Dirac representation $\Gamma_\varepsilon$ is a real matrix:
\begin{equation}\label{gs}
\Gamma_+=\left(\begin{array}{cccc}0&0&2&0\\ 0&0&0&0\\0&0&0&0\\0&2&0&0
\end{array}\right),\qquad \Gamma_-=-\left(\begin{array}{cccc}0&0&0&0\\ 0&0&0&2\\2&0&0&0\\0&0&0&0
\end{array}\right).
\end{equation}
With this forms of the matrix $\Gamma$, equation (\ref{diraceq2p}) maintains its Lorentz covariant nature and invariance with respect to reflections. It can be verified that all the components of the bispinor  solutions  satisfy (\ref{pareq}). In the case of massless fermions the two signs of $\varepsilon$ are related to the helicity transformation applied to $\widetilde{\psi}(\mathbf{r})$.

\subsection{Solutions}\label{sopa}

It would be valuable to find the explicit solutions of the paraxial Dirac equation which has been postulated in the preceding subsection, and to compare them to those derived in Sect.~\ref{potb} via superpositions of exact modes weighted with certain appropriately chosen Gaussian factors. 

Eq. (\ref{diraceq2p}) rewritten in cylindrical coordinates reads:
\begin{equation}\label{diraceq2pa}
\left(\gamma^0E+i\gamma^\rho\partial_\rho+\frac{i}{\rho}\,\gamma^\varphi\partial_\varphi+i\Gamma\partial_z-\gamma^3p -m\right)\widetilde{\psi}(\mathbf{r})=0.
\end{equation}
Its solutions can be constructed in a simple way. Consider any scalar function $\widetilde{f}_n(\rho,\varphi,z)$ satisfying the equation 
\begin{equation}\label{eqpar}
\left[\partial_\rho^2+\frac{1}{\rho}\,\partial _\rho+\frac{1}{\rho^2}\,\partial_\varphi^2+2ip\partial_z\right]\widetilde{f}_n(\rho,\varphi,z)=0,
\end{equation}
which will be specified below. A number of such functions are known in optics. Then one can subsitute
\begin{equation}\label{biep}
\widetilde{\psi}_{n\uparrow \downarrow}(\rho,\varphi,z)=\left[\begin{array}{c}\widetilde{\phi}_{n\uparrow \downarrow}(\rho,\varphi,z)\\ \widetilde{\chi}_{n\uparrow \downarrow}(\rho,\varphi,z)\end{array}\right],
\end{equation}
where
\begin{equation}\label{spu}
\widetilde{\phi}_{n\uparrow}=\left(\begin{array}{c}\widetilde{f}_n\\ 0\end{array}\right)
\end{equation}
for ``spin-up'', and 
\begin{equation}\label{spd}
\widetilde{\phi}_{n\downarrow}=\left(\begin{array}{c}0 \\ \widetilde{f}_n\end{array}\right)
\end{equation}
for ``spin-down'' solutions. After plugging these expressions into (\ref{diraceq2pa}), the equation for the lower bispinor components is derived in the form of
\begin{eqnarray}
\widetilde{\chi}_{n\uparrow \downarrow}&=&\frac{1}{E+m}\Big[-i\sigma_\rho\partial_\rho+\frac{n}{\rho}\,\sigma_\varphi\label{chud}\\
&&-i(\sigma_z-\varepsilon\openone)\partial_z+\sigma_z p\Big]\widetilde{\phi}_{n\uparrow\downarrow},\nonumber
\end{eqnarray} 
and hence
\begin{equation}\label{genuu}
\widetilde{\psi}_{n\uparrow}(\rho,\varphi,z)= \left[\begin{array}{c}\widetilde{f}_n\\ 0\\ \frac{1}{E+m}\left(\begin{array}{c}p\widetilde{f}_n-i(1-\varepsilon)\partial_z\widetilde{f}_n\\ ie^{i\varphi}(-\partial_\rho\widetilde{f}_n+\frac{n}{\rho}\widetilde{f}_n)\end{array}\right)\end{array}\right],
\end{equation}
\begin{equation}\label{gendd}
\widetilde{\psi}_{n\downarrow}(\rho,\varphi,z)= \left[\begin{array}{c}0\\ \widetilde{f}_n\\ \frac{1}{E+m}\left(\begin{array}{c}-ie^{-i\varphi}(\partial_\rho\widetilde{f}_n+\frac{n}{\rho}\widetilde{f}_n) \\ -p\widetilde{f}_n+i(1+\varepsilon)\partial_z\widetilde{f}_n \end{array}\right)\end{array}\right],
\end{equation}
correspondingly.

In the following, the four different scalar functions $\widetilde{f}_n(\rho,\varphi,z)$ will be substituted into (\ref{genuu}) and (\ref{gendd}) in order to obtain the solutions in the form of Gaussian, Bessel-Gaussian, modified Bessel-Gaussian and elegant Laguerre-Gaussian beams.

\subsubsection{Gaussian beam}\label{gabe}

For the pure Gaussian beam one chooses the fundamental solution of (\ref{eqpar}) in the form
\begin{equation}\label{gaug}
\widetilde{f}_n(\rho,\varphi,z)=e^{in\varphi}\left(\frac{2}{\alpha(z)}\right)^{n+1}\!\!\!\!\rho^ne^{-\frac{\rho^2}{\alpha(z)}}.
\end{equation}
With this formula it is straightforward to show that
\begin{subequations}\label{fnp}
\begin{align}
ie^{i\varphi}\left(-\partial_\rho\widetilde{f}_n+\frac{n}{\rho}\widetilde{f}_n\right)&=i\widetilde{f}_{n+1},\label{fnp1}\\
\intertext{and}
-ie^{-i\varphi}\left(\partial_\rho\widetilde{f}_n+\frac{n}{\rho}\widetilde{f}_n\right)&=2p\partial_z\widetilde{f}_{n-1}.\label{fnp2}
\end{align}
\end{subequations}
Consequently the following solutions for the Gaussian beam of Dirac particles can be written down:
\begin{equation}\label{tgapu}
\widetilde{\psi}_{n\uparrow}(\rho,\varphi,z)= \left[\begin{array}{c}\widetilde{f}_n\\ 0\\ \frac{1}{E+m}\left(\begin{array}{c}p\widetilde{f}_n-i(1-\varepsilon)\partial_z\widetilde{f}_n\\ i\widetilde{f}_{n+1}\end{array}\right)\end{array}\right],
\end{equation}
for ``spin-up'' modes, and
\begin{equation}\label{tgapd}
\widetilde{\psi}_{n\downarrow}(\rho,\varphi,z)= \left[\begin{array}{c}0\\ \widetilde{f}_n\\ \frac{1}{E+m}\left(\begin{array}{c}2p\partial_z\widetilde{f}_{n-1}\\ -p\widetilde{f}_n+i(1+\varepsilon)\partial_z\widetilde{f}_n \end{array}\right)\end{array}\right].
\end{equation}
for ``spin-down'' ones. These formulas are to be compared to (\ref{gapu}) and (\ref{gapd}) obtained via the GPT of Sect. \ref{gaud}, i.e., through the optical-like superpositions of exact (nonparaxial) modes. Both beams turn out to be identical up to terms negligible within the paraxial approximation [cf. (\ref{apl})]. The particular choice of the value of the parameter $\varepsilon$ (i.e. $+1$ or $-1$) is inessential since $\varepsilon\partial_z \widetilde{f}$ is always accompanied by $p\widetilde{f}$ and
\begin{equation}\label{}
 p\widetilde{f}+\varepsilon \partial_z \widetilde{f}=p (\widetilde{f}+\varepsilon\lambda_{\mathrm{dB}}\partial_z \widetilde{f})\approx  p\widetilde{f},
\end{equation}
where $\lambda_{\mathrm{dB}}$ stands for de Broglie wavelength of the Dirac particle.

\subsubsection{Bessel-Gaussian beam}\label{begabe}

In order to generate a paraxial beam that has the Bessel-Gaussian characteristics, one needs to choose the following solution of (\ref{eqpar}):
\begin{equation}\label{gagb}
\widetilde{f}_n(\rho,\varphi,z)=e^{in\varphi}\,\frac{2}{\alpha(z)}\,e^{\frac{\chi^2-\rho^2}{\alpha(z)}}J_n\left(\frac{2\chi\rho}{\alpha(z)}\right),
\end{equation}
with some parameter $\chi$. Now, using identities (\ref{ide}) one arrives at
\begin{subequations}\label{bfnp}
\begin{align}
ie^{i\varphi}\left(-\partial_\rho\widetilde{f}_n+\frac{n}{\rho}\widetilde{f}_n\right)&=\frac{2i\rho}{\alpha(z)}\,e^{i\varphi}\widetilde{f}_n+\frac{2i\chi}{\alpha(z)}\,\widetilde{f}_{n+1},\label{bfnp1}\\
\intertext{and}
-ie^{-i\varphi}\left(\partial_\rho\widetilde{f}_n+\frac{n}{\rho}\widetilde{f}_n\right)&=\frac{2i\rho}{\alpha(z)}\,e^{-i\varphi}\widetilde{f}_n-\frac{2i\chi}{\alpha(z)}\,\widetilde{f}_{n-1},\label{bfnp2}
\end{align}
\end{subequations}
which allows to write down the formulas for both paraxial solutions:
\begin{equation}\label{tbgapu}
\widetilde{\psi}_{n\uparrow}(\rho,\varphi,z)= \left[\begin{array}{c}\widetilde{f}_n\\ 0\\ \frac{1}{E+m}\left(\begin{array}{c}p\widetilde{f}_n-i(1-\varepsilon)\partial_z\widetilde{f}_n\\ \frac{2i\rho}{\alpha(z)}\,e^{i\varphi}\widetilde{f}_n+\frac{2i\chi}{\alpha(z)}\,\widetilde{f}_{n+1}\end{array}\right)\end{array}\right],
\end{equation}
\begin{equation}\label{tbgapd}
\widetilde{\psi}_{n\downarrow}(\rho,\varphi,z)= \left[\begin{array}{c}0\\ \widetilde{f}_n\\ \frac{1}{E+m}\left(\begin{array}{c}\frac{2i\rho}{\alpha(z)}\,e^{-i\varphi}\widetilde{f}_n-\frac{2i\chi}{\alpha(z)}\,\widetilde{f}_{n-1}\\ -p\widetilde{f}_n+i(1+\varepsilon)\partial_z\widetilde{f}_n \end{array}\right)\end{array}\right].
\end{equation}
Up to neglibible terms these expressions are identical to (\ref{bgapu}) and (\ref{bgapd}) respectively.

\subsubsection{Modified Bessel-Gaussian beam}\label{mbegabe}

The same consistency is obtained for modified Bessel-Gaussian beams, upon choosing
\begin{equation}\label{fobd}
\widetilde{f}_n(\rho,\varphi,z)=e^{in\varphi}\,\frac{2}{\alpha(z)}\,e^{-\frac{\chi^2+\rho^2}{\alpha(z)}}I_n\left(\frac{2\chi\rho}{\alpha(z)}\right).
\end{equation}
The identities for the hyperbolic Bessel functions
\begin{subequations}\label{jde}
\begin{align}
&x\,\frac{d}{dx}I_n(x)-nI_n(x)=xI_{n+1}(x),\label{jde1}\\
&x\,\frac{d}{dx}I_n(x)+nI_n(x)=xI_{n-1}(x),\label{jde2}
\end{align}
\end{subequations}
imply
\begin{subequations}\label{bfnpa}
\begin{align}
ie^{i\varphi}\left(-\partial_\rho\widetilde{f}_n+\frac{n}{\rho}\widetilde{f}_n\right)&=\frac{2i\rho}{\alpha(z)}\,e^{i\varphi}\widetilde{f}_n-\frac{2i\chi}{\alpha(z)}\,\widetilde{f}_{n+1},\label{bfnpa1}\\
-ie^{-i\varphi}\left(\partial_\rho\widetilde{f}_n+\frac{n}{\rho}\widetilde{f}_n\right)&=\frac{2i\rho}{\alpha(z)}\,e^{-i\varphi}\widetilde{f}_n-\frac{2i\chi}{\alpha(z)}\,\widetilde{f}_{n-1},\label{bfnpa2}
\end{align}
\end{subequations}
and finally
\begin{equation}\label{mtbgapu}
\widetilde{\psi}_{n\uparrow}(\rho,\varphi,z)= \left[\begin{array}{c}\widetilde{f}_n\\ 0\\ \frac{1}{E+m}\left(\begin{array}{c}p\widetilde{f}_n-i(1-\varepsilon)\partial_z\widetilde{f}_n\\ \frac{2i\rho}{\alpha(z)}\,e^{i\varphi}\widetilde{f}_n-\frac{2i\chi}{\alpha(z)}\,\widetilde{f}_{n+1}\end{array}\right)\end{array}\right],
\end{equation}
\begin{equation}\label{mtgapd}
\widetilde{\psi}_{n\downarrow}(\rho,\varphi,z)= \left[\begin{array}{c}0\\ \widetilde{f}_n\\ \frac{1}{E+m}\left(\begin{array}{c}\frac{2i\rho}{\alpha(z)}\,e^{-i\varphi}\widetilde{f}_n-\frac{2i\chi}{\alpha(z)}\,\widetilde{f}_{n-1}\\ -p\widetilde{f}_n+i(1+\varepsilon)\partial_z\widetilde{f}_n \end{array}\right)\end{array}\right],
\end{equation}
in agreement with (\ref{mbgapu}) and (\ref{mbgapd}).

\subsubsection{Elegant Laguerre-Gaussian beam}\label{elegabe}

The last beam dealt with in the previous section was the eLG beam, for which one chooses
\begin{eqnarray}
\widetilde{f}_{n,q}(\rho,\varphi,z)&=&2^qq!\,e^{in\varphi}\left(\frac{2}{\alpha(z)}\right)^{n+q+1}\nonumber\\ &&\times\rho^n\,e^{-\frac{\rho^2}{\alpha(z)}}L_q^{(n)}\left(\frac{\rho^2}{\alpha(z)}\right).\label{folg}
\end{eqnarray}
Using the identity \cite{ospa}
\begin{subequations}\label{lsge}
\begin{align}
&L_q^{(n)}(x)-\frac{d}{dx}L_q^{(n)}(x)=L_q^{(n)}(x)-L_{q-1}^{(n+1)}(x)=L_q^{(n+1)}(x),\label{lsge1}\\
\intertext{and}
&(n-x)L_q^{(n)}(x)+x\,\frac{d}{dx}L_q^{(n)}(x)=(q+1)L_{q+1}^{(n-1)}(x),\label{lsge2}
\end{align}
\end{subequations}
where the latter can be easily proved with the use of the Rodrigues formula
\begin{equation}\label{rodri}
L_q^{(n)}(x)=\frac{e^x x^{-n}}{q!}\,\frac{d^q}{dx^q}\left(e^{-x}x^{q+n}\right),
\end{equation}
one can show, that
\begin{subequations}\label{lgn}
\begin{align}
ie^{i\varphi}\left(-\partial_\rho\widetilde{f}_{n,q}+\frac{n}{\rho}\widetilde{f}_{n,q}\right)&=i\widetilde{f}_{n+1,q},\label{lgn1}\\
-ie^{-i\varphi}\left(\partial_\rho\widetilde{f}_{n,q}+\frac{n}{\rho}\widetilde{f}_{n,q}\right)&=-i\widetilde{f}_{n-1,q+1}.\label{lgn2}
\end{align}
\end{subequations}
This leads to the following expressions for paraxial ``spin-up'' and ``spin-down'' states:
\begin{equation}\label{lgud}
\widetilde{\psi}_{n,q\uparrow}(\rho,\varphi,z)= \left[\begin{array}{c}\widetilde{f}_{n,q}\\ 0\\ \frac{1}{E+m}\left(\begin{array}{c}p\widetilde{f}_{n,q}-i(1-\varepsilon)\partial_z\widetilde{f}_{n,q}\\ i\,\widetilde{f}_{n+1,q}\end{array}\right)\end{array}\right],
\end{equation}
\begin{equation}\label{lgdd}
\widetilde{\psi}_{n,q\downarrow}(\rho,\varphi,z)= \left[\begin{array}{c}0\\ \widetilde{f}_{n,q}\\ \frac{1}{E+m}\left(\begin{array}{c}-i\widetilde{f}_{n-1,q+1}\\ -p\widetilde{f}_{n,q}+i(1+\varepsilon)\partial_z\widetilde{f}_{n,q} \end{array}\right)\end{array}\right],
\end{equation}
in accordance with (\ref{elgapu}) and (\ref{elgapd}).

\subsection{Comparison to Foldy-Wouthuysen paraxial beams}\label{fw}

It is interesting to confront these paraxial beams to those obtained in the well-established way, which exploits the Foldy-Wouthuysen representation \cite{folw}. It is known that in the case of a free Dirac particle it is possible to entirely decouple the upper (large) bispinor components from the lower (small) ones, by performing the unitary transformation
\begin{equation}\label{fwtr}
\Psi_{\mathrm{FW}}=e^{i\hat{S}}\Psi,
\end{equation}
where 
\begin{equation}\label{est}
\hat{S}=-i\hat{\theta}\frac{\bm{\gamma\,\hat{p}}}{\hat{p}},\qquad \tan 2\hat{\theta} =\frac{\hat{p}}{m}.
\end{equation}
The ``hats'' over the symbols are placed in order to stress their operatorial character.
Then, for a definite spin, the particle state is characterized by one scalar function for which, in the paraxial regime, the standard equation (\ref{eqpar}) is obtained. Apart from the common constant coefficient, this function plays then the role analogous to $e^{ipz}\widetilde{f}_n(\rho,\varphi,z)$. 

Now, in order to compare the FW paraxial approximation to ours, it is necessary to invert the transformation (\ref{fwtr}) and to recover the entire Dirac bispinor. It cannot be done in an explicit way due to the nonlocal character of the FW transformation. However, within the paraxial approximation, owing to the condition $\partial_z \widetilde{f}_n(\rho,\varphi,z)\ll p \widetilde{f}_n(\rho,\varphi,z)$, it can be shown that both approaches lead to the same expressions.

In cylindrical coordinates the operator $e^{-i\hat{S}}$ may be given the form
\begin{eqnarray}
e^{-i\hat{S}}=&&\frac{1}{\sqrt{2\hat{E}(\hat{E_p}+m)}}\label{opems}\\
&&\left(m+\hat{E}+i\gamma^\rho\partial_\rho+\frac{i}{\rho}\,\gamma^{\varphi}\partial_\varphi+i\gamma^z\partial_z\right),\nonumber
\end{eqnarray}
where $\hat{E}=\sqrt{m^2+\hat{p}^2}$. This operator 
acts on the bispinor
\begin{equation}\label{bifw}
\Psi_{FW}=e^{ipz}\sqrt{\frac{2E}{E+m}}\left[\begin{array}{c}\widetilde{f}_{n}\\ 0\\ 0\\ 0\end{array}\right],
\end{equation}
where $\widetilde{f}_{n}$ satisfies the scalar paraxial equation.

The nonlocal operators $\hat{E_p}$ acting on $e^{ipz}\widetilde{f}_n(\rho,\varphi,z)$, can be expanded in terms of powers of $\hat{p}$, and then, as told above, all small contributions are omitted. This resuts in the simple replacement: $\hat{E}\mapsto E$, and one gets  
\begin{equation}\label{fgapu}
\widetilde{\psi}_{n\uparrow}(\rho,\varphi,z)= \left[\begin{array}{c}\widetilde{f}_n\\ 0\\ \frac{1}{E+m}\left(\begin{array}{c}p\widetilde{f}_n-i\partial_z\widetilde{f}_n\\ ie^{i\varphi}(-\partial_\rho\widetilde{f}_n+\frac{n}{\rho}\, \widetilde{f}_n)\end{array}\right)\end{array}\right],
\end{equation}
in agreement with (\ref{genuu}). Similar results can be obtained for ``spin-down'' solution.

\section{Summary}\label{sum}

In this work, the paraxial Dirac equation is proposed, in which the matrix $\gamma^5$ additionally occurs accompanying the derivative along the propagation axis. It is shown that this ensures each of the four components of the bispinor to satisfy the scalar paraxial equation well known from optics.

The upper components of the four solutions obtained from this equation, exhibit the form  typical for corresponding scalar optical beams carrying orbital angular momentum and, therefore, belong to the family of vortex beams. The lower components are described by functions for which the angular momentum differs by $\pm \hbar$. This implies that these beams can be eigenstates of neither orbital angular momentum nor spin.

In the proposed equation the additional parameter $\varepsilon$ occurs, which can assume two values: $\pm 1$. The specific choice of these only affects those terms in the lower components of the bispinor, which can be omitted anyway within the framework of the paraxial approximation. 
Still under this approximation, the obtained functions conform also to those found earlier by the use of the Foldy-Wouthuysen representation.

It would be interesting to proceed beyond the basic paraxial approximation and to obtain corrections to the fundamental solutions, which could reveal the physical meaning of the parameter $\varepsilon$. In this context, it would be worthwhile to specify certain ``paraxial'' transformation that, when applied in a systematic way to the Dirac equation, would enable one to obtain successive approximations, similar to the Foldy transformation in the electromagnetic field.

In Section \ref{potb}, a different approach is applied. Certain integral transformations, denoted successively as GPT, BGPT, mBGPT and eLGPT, performed on the rigorous Dirac solution in cylindrical coordinates are defined, leading to the same four paraxial beams. These integral transformations correspond to superpositions of exact, but non-physical modes with some particular weighting factors. Both approaches are shown to agree with each other up to expressions corresponding to the ratio of the electron de Broglie wavelength to the spatial dimension of the beam.

\end{document}